\documentclass[superscriptaddress,aps,prb,reprint,doublecolumn,notitlepage,floatfix]{revtex4-1}
\usepackage{graphicx}
\usepackage{epstopdf}
\usepackage{braket}
\usepackage{mathrsfs}
\usepackage{amsmath}
\newcommand{\crmn}{\text{Cr}_7\text{Mn}}

\begin{document}

\title{Constructing Clock-Transition-Based Two-Qubit Gates from Dimers of Molecular Nanomagnets}

\author{Charles A. Collett}
\affiliation{Department of Physics and Astronomy, Amherst College, Amherst, MA 01002}
\affiliation{Department of Physics, Hamilton College, Clinton, NY 13323}
\affiliation{Department of Physics, Muhlenberg College, Allentown, PA 18104}
\author{Paolo Santini}
\author{Stefano Carretta}
\affiliation{Dipartimento di Fisica e Scienze della Terra, Universit\`{a} di Parma, Parma 43123, Italy}
\author{Jonathan R. Friedman}
\email[Corresponding author: ]{jrfriedman@amherst.edu}
\affiliation{Department of Physics and Astronomy, Amherst College, Amherst, MA 01002}
\date{\today}

\begin{abstract}
A good qubit must have a coherence time long enough for gate operations to be performed.  Avoided level crossings allow for clock transitions in which coherence is enhanced by the insensitivity of the transition to fluctuations in external fields.  Because of this insensitivity, it is not obvious how to effectively couple qubits together while retaining clock-transition behavior.  Here we present a scheme for using a heterodimer of two coupled molecular nanomagnets, each with a clock transition at zero magnetic field, in which all of the gate operations needed to implement one- and two-qubit gates can be implemented with pulsed radio-frequency radiation.  We show that given realistic coupling strengths between the nanomagnets in the dimer, good gate fidelities ($\sim$99.4\%) can be achieved.  We identify the primary sources of error in implementing gates and discuss how these may be mitigated, and investigate the range of coherence times necessary for such a system to be a viable platform for implementing quantum computing protocols.  
\end{abstract}

\maketitle

A variety of physical systems have been explored as possible qubits~\cite{ladd_quantum_2010}, including superconducting devices~\cite{devoret_superconducting_2013}, trapped ions~\cite{blatt_entangled_2008}, and both electronic and nuclear spin systems~\cite{loss_quantum_1998,nielsen_quantum_2010,friedman_single-molecule_2010,dutt_quantum_2007}. The ideal multi-qubit architecture would have an array of independently controlled, long-lived qubits, with adjustable couplings between each pair of qubits. Physical implementations of qubits involve trade offs between various important features, such as coherence times, addressability, and scalability. Electronic spin systems have garnered a fair amount of attention in recent years as potential qubits~\cite{Jenkinsscalablearchitecturequantum2016}, especially in the context of hybrid quantum architectures, in which spins could fulfill the role of memory qubits~\cite{KuboHybridQuantumCircuit2011,KuboStrongCouplingSpin2010}.


Molecule-based spin systems, such as molecular nanomagnets (MNMs), offer several advantages over other types of spin systems. In particular, because they are chemically synthesized, properties such as the spin Hamiltonian and interactions with environmental degrees of freedom can be chemically engineered. A class of heterometallic rings has been extensively studied as possible qubits~\cite{affronte_ring_2007}. One of the most-studied of these are the family Cr$_7$M, where M is a transition-metal ion~\cite{larsen_synthesis_2003,collettClockTransitionCr7Mn2019}. These systems offer the ability to engineer the total ground-state spin of the system through choice of M. A combination of dilution of the molecules in a non-magnetic medium and chemical engineering by using different ligands and cations to maximize coherence has yielded $T_2\sim15$~$\mu$s~\cite{wedge_chemical_2012}.

\begin{figure}[ht!]
\includegraphics[width=.9\linewidth]{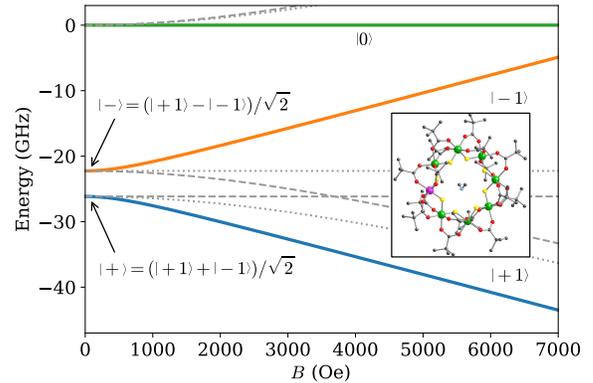}
\caption{\label{fig:monomer}Energy level diagram for a single $\crmn$ molecule, showing the zero field avoided crossing between $\ket{m=\pm1}$ states, creating the $\ket{\pm}$ clock states. Solid lines show dependence on a field along the easy-axis (z) direction. Dashed (dotted) lines correspond to the dependence on field along the hard (medium) axis. Inset: Molecular structure of $\crmn$.}
\end{figure}

An important source of decoherence in many qubit spin systems comes about from fluctuations in local electromagnetic fields that change the spin's energy (by, e.g., the Zeeman effect) and thereby induce fluctuations in the phase of the spin's quantum state.  One effective technique to ameliorate this mechanism of decoherence is to make use of so-called atomic-clock transitions in which the energy levels of a qubit depend non-linearly on the field in some region (i.e.~near an avoided level crossing). In particular, when the transition frequency between levels is independent of field at some field ($df/dB=0$), the transition will be immune to field fluctuations to first order, suppressing decoherence from those fluctuations and concomitantly increasing the coherence time $T_2$.  This technique has been exploited with great effect in superconducting qubits, where it is often referred to as the ``sweet spot"~\cite{vion_manipulating_2002}. Recently, the implementation of clock transitions in Ho-based~\cite{shiddiq_enhancing_2016} and Co-based~\cite{ZadroznyPorousArrayClock2017} MNM systems has produced a marked enhancement of $T_2$ in the vicinity of avoided crossings, resulting in $T_2$ as high as $\sim8~\mu$s and $\sim14~\mu$s, respectively. 
Similarly, clock-transition behavior has been observed in the heterometallic ring $\crmn$ ($S=1$), discussed below~\cite{collettClockTransitionCr7Mn2019}.
Such rings can be coupled to each other to form supramolecular dimers, and possibly longer chains, that can be used to build multiqubit systems with  coherence times comparable to those of the constituent monomers~\cite{chiesa_molecular_2015,ferrando-soria_modular_2016,timco_heterodimers_2016,timco_engineering_2009,timco_linking_2011,ardavan_engineering_2015,whiteheadRingsThreadsLinkers2013}. 

Here we describe a scheme in which MNMs displaying clock transitions (such as $\crmn$) can be joined into dimers in which the resulting states of the coupled system retain the characteristics of clock transitions.  Remarkably, although in our scheme the molecular monomers exchange couple to each other, they remain insensitive to field fluctuations, thus enabling single- and two-qubit operations to be implemented while preserving the immunity of the system to field fluctuations. Recent work on dimers of Ti atoms has shown the efficacy of clock transitions in reducing decoherence in dimer systems with exchange coupling~\cite{bae_enhanced_2018,vorndamme_decoherence_2020}, though that work involved a single two-state clock transition.  In contrast, the dimers described herein present a manifold of four states connected by clock transitions, thus providing a two-qubit system. 

An effective Hamiltonian for an isolated $S=1$ $\crmn$ molecule is
\begin{equation}
\mathscr{H}_i=-D_iS_{iz}^2 + E_i(S_{ix}^2-S_{iy}^2)+g_i \mu_B \vec{S}_i\cdot\vec{B}, \label{eq:Hdimer_0}
\end{equation}
The $D_i$ term represents the system's axial (easy-axis) anisotropy, while the $E_i$ term corresponds to the transverse anisotropy. Here the subscript $i$ designates a particular molecule. Such a Hamiltonian can be justified as the low-energy approximation resulting from an \emph{ab initio} treatment of the ring~\cite{chiesaManybodyInitioStudy2016,chiesaManyBodyModelsMolecular2013}.  In addition, numerous experimental results confirm the validity of this effective Hamiltonian at low temperatures~\cite{ardavan_will_2007,collettClockTransitionCr7Mn2019}.  We can identify the $S_{zi}$ eigenstates by their $m$ value: $\ket{m=0}\equiv\ket{0}$ and $\ket{m=\pm1}\equiv\ket{\pm1}$. At zero field, the energy eigenstates are $\ket{0}$ and $\ket{\pm}=\left(\ket{+1}\pm\ket{-1}\right)/\sqrt{2}$.  The latter two states exhibit an avoided crossing with a ``tunnel splitting" of $2E_i$.  Figure \ref{fig:monomer} shows the energy eigenstates for this system as a function of field applied along the easy (z), medium (y) and hard (x) axial directions.  The figure illustrates that the zero-field transition between the two lowest-energy states is independent to first order to any component of the magnetic field and thus constitutes an atomic-clock transition, with a significant transition matrix element for the $S_z$ operator: $\bra{+}S_z\ket{-}=1$. Through variations in synthesis, molecules with different values of parameters ($D_i$ and $E_i$) can be produced, notably the so-called green and purple variants of $\crmn$~\cite{timco_linking_2011}.

When coupled together, a pair of molecules with different parameters form a supramolecular heterodimer~\cite{timco_heterodimers_2016}.   Interactions between the spins in the dimer can be modeled as a bilinear exchange interaction:
\begin{equation}
\mathscr{H}_{\rm{J}}=\vec{S}_1\cdot J\cdot\vec{S}_2= \vec{S}_1\cdot\tilde{J}\cdot\vec{S}_2+J_{zz} S_{1z}S_{2z} .\label{eq:Hdimer_int}
\end{equation}
We isolate the  $J_{zz} $ term here (and implicitly define the $\tilde{J}$ tensor) because it is the only term that  directly couples any of the four lowest-energy states to each other.   As a consequence, this term is responsible for an error in the implementation of single-qubit rotations, as will be discussed below.  It is important to note that molecules within the dimer need not have any simple relative orientation and, thus, each of the principal (easy, medium and hard) axes of the two spins may have any relative orientation.  As a consequence, the components of $J$ do not necessarily refer to specific directions in space but to couplings between different axial directions of each spin; e.g. $J_{xz}$ describes the coupling between the hard-axis component of spin 1 and the easy-axis component of spin 2.

\begin{figure}[ht!]
\includegraphics[width=.9\linewidth]{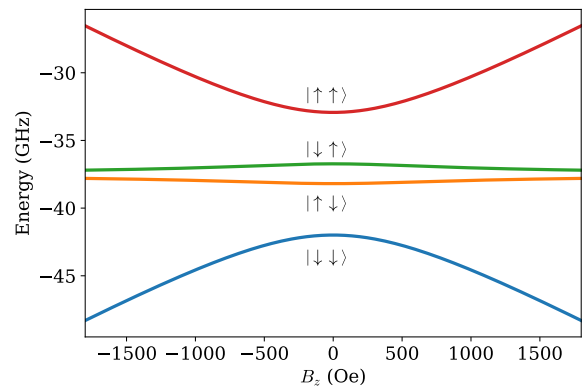}
\caption{\label{fig:dimer_levels} Energies vs.~field for the four lowest states of a heterodimer of $\crmn$ molecules calculated using Eq.~\ref{HJ2} with $J_{zz}=$~50~MHz and $J_{\perp}=$~100~MHz, demonstrating the retention of clock states in the coupled system.  For clarity, only the dependence on the z component of magnetic field is shown.}
\end{figure}

The total zero-field Hamiltonian for the system is 
\begin{equation}
\mathscr{H}=\mathscr{H}_1+\mathscr{H}_2+\mathscr{H}_{\rm{J}}. \label{eq:Hdimer_tot}
\end{equation}
When the $D_i$ are much larger than all the other energy parameters $\left(E_i, J_{ij}\right)$, 
the subspace of the four lowest-energy states acts as a system of two coupled effective $S=1/2$ spins. 
For $\tilde{J}=0$ and the realistic case of $E_i\gg J_{zz}$, the lowest and highest energy states in the subspace are to a good approximation $\ket{++}$ and $\ket{--}$, with energies $\mathscr{E}_{\pm\pm}=2\left(-\bar{D}\pm\bar{E}\right)$, where $\bar{D}=\left(D_1+D_2\right)/2$ and $\bar{E}=\left(E_1+E_2\right)/2$.  The two middle-energy states can be represented as
\begin{eqnarray}
\ket{\uparrow \downarrow}&=\cos\theta\ket{+-}+\sin\theta\ket{-+},\nonumber\\
\ket{\downarrow \uparrow}&=-\sin\theta\ket{+-}+\cos\theta\ket{-+},
\end{eqnarray}
where $\tan{2\theta}=\frac{2J_{zz}}{\Delta E}$, with energies $\mathscr{E}_{\substack{\uparrow\downarrow\\ \downarrow\uparrow}}=-2\bar{D}\pm\sqrt{\Delta E^2+J_{zz}^2}$, respectively, defining $\Delta E = E_1 - E_2$. Since the states are constructed from clock states, near zero field all four of these states are barely affected by a magnetic field along any direction, as illustrated for the z component of field in Fig.~\ref{fig:dimer_levels}, unlike real coupled $S = 1/2$ spins.  For implementation of quantum-computing protocols we use the energy eigenstates as the logical basis, labelling these with vertical arrows, e.g.~$\ket{\uparrow \downarrow}$.

Certain transitions within the four-state manifold are degenerate, \textit{e.g.} $\ket{++}\leftrightarrow\ket{\uparrow \downarrow}$ is degenerate with $\ket{\downarrow\uparrow }\leftrightarrow\ket{--}$. These degeneracies are broken by the  $\tilde{J}$ term in 
Eq.~(\ref{eq:Hdimer_int}).  For simplicity, we consider the case in which $\tilde{J}$ is diagonal such that 
\begin{equation}
\label{HJ2}
\mathscr{H}_J= J_{\perp}(S_{1x}S_{2x}+S_{1y}S_{2y})+J_{zz}S_{1z}S_{2z}
\end{equation}
(Other forms of $\tilde{J}$ give qualitatively similar results.) 
To second order in $J_\perp$,   
the $\ket{++}$ and $\ket{--}$ states become, respectively, the states $\ket{\uparrow\uparrow}$ and 
$\ket{\downarrow\downarrow}$:
\begin{multline}
    \label{psi++}
    \ket{\substack{\uparrow\uparrow\\ \downarrow\downarrow}}=\left(1-\frac{J_\perp^2}{2\mathcal{E}_{\pm\pm}^2}\right)\ket{\pm\pm}+\frac{J_\perp}{\mathcal{E}_{\pm\pm}}\ket{00}\\
    \pm\frac{J_\perp^2}{\left(\mathcal{E}_{--}-\mathcal{E}_{++}\right)\mathcal{E}_{\pm\pm}}\ket{\mp\mp}
\end{multline}
and their energies are shifted by $\Delta \mathscr{E} =\frac{J_\perp^2}{2\left(-\bar{D}\pm\bar{E}\right)}$, respectively, thereby breaking the transition degeneracies.  The $\ket{\uparrow\downarrow}$ and $\ket{\downarrow\uparrow}$ states are unchanged by  $\tilde{J}$.  Under these circumstances, the four-state system becomes an effective two-qubit system with constant coupling in which standard one- and two-qubit gates can be implemented with pulsed radiation.

Transitions between the four states are induced by radio-frequency (rf) radiation:
\begin{equation}
\mathscr{H}_{\rm{rf}}=g\mu_B \vec{B}_{rf}\cdot\left(\vec{S}_{1}+\vec{S}_{2}\right).
\label{eq:Hrf}
\end{equation}
We consider a radiation field $\vec{B}_{rf}=\vec{B}_1\cos\left(\omega_1 t+\phi_1\right)+\vec{B}_2\cos\left(\omega_2 t+\phi_2\right)$ consisting of up to two radiation frequencies $\omega_1$ and $\omega_2$ with amplitudes $\vec{B}_1$ and $\vec{B}_2$, respectively. In our calculations, $\omega_1$ is set to match the average frequency of the  $\ket{\downarrow\downarrow}\rightarrow\ket{\downarrow\uparrow}$ and $\ket{\uparrow\downarrow}\rightarrow\ket{\uparrow\uparrow}$ transitions while $\omega_2$ matches the average for the $\ket{\downarrow\downarrow}\rightarrow\ket{\uparrow\downarrow}$ and $\ket{\downarrow\uparrow}\rightarrow\ket{\uparrow\uparrow}$ transitions.  Only the z component of $\vec S_i$ provides non-zero matrix elements for these transitions, meaning that the radiation coupling can be reduced to  
\begin{equation}
\mathscr{H}_{\rm{rf}}=g\mu_B \left(B_{1z}S_{1z}+B_{2z}{S}_{2z}\right),
\label{eq:Hrf_reduced}
\end{equation}
dropping terms that correspond to far-off-resonance transitions.  We note that since the easy axes of the two spins are not in general parallel, the z components of the radiation fields may correspond to different directions even if $\vec{B}_1$ and $\vec{B}_2$ are colinear.   

We simulated our system by solving the Schr\"odinger equation to find the time evolution under the Hamiltonian $\mathscr{H}+\mathscr{H}_{\rm{rf}}$. The Hamiltonian was transformed into the interaction picture using the operator $U_{\rm{int}}=e^{-i\mathscr{\tilde H}t}$, where 

\begin{multline}
\mathscr{\tilde H}=\hbar\omega_1\ket{\downarrow\uparrow}\bra{\downarrow\uparrow}+\hbar\omega_2\ket{\uparrow\downarrow}\bra{\uparrow\downarrow}\\+\left(\hbar{\omega_1}+\hbar{\omega_2}\right)\ket{\uparrow\uparrow}\bra{\uparrow\uparrow},
\end{multline}

After discarding rapidly oscillating terms in the Hamiltonian (rotating wave approximation) as well as dropping an irrelevant constant, one obtains the interaction-picture effective Hamiltonian:

\begin{multline}
\mathscr{H}_{\rm{int}}=\delta\left(\ket{\downarrow\uparrow}\bra{\downarrow\uparrow}+\ket{\uparrow\downarrow}\bra{\uparrow\downarrow}\right)+\frac{g\mu_B}{2}\times\\
[(B_1 e^{i\phi_1}\left(S_{1z,\downarrow\downarrow,\downarrow\uparrow}\ket{{\downarrow\downarrow}}\bra{\downarrow\uparrow}+S_{1z,\uparrow\downarrow,\uparrow\uparrow}\ket{\uparrow\downarrow}\bra{\uparrow\uparrow}\right)\\
+B_2 e^{i\phi_2}\left(S_{2z,\downarrow\downarrow,\uparrow\downarrow}\ket{\downarrow\downarrow}\bra{\uparrow\downarrow}+S_{2z,\downarrow\uparrow,\uparrow\uparrow}\ket{\downarrow\uparrow}\bra{\uparrow\uparrow}\right)\\+h.c.],
\end{multline}
where $\delta=-\frac{2\bar D J_\perp^2}{\bar D^2-\bar E^2}$ and $S_{1z,\downarrow\downarrow,\downarrow\uparrow}=\bra{{\downarrow\downarrow}}S_{1z}\ket{\downarrow\uparrow}$, etc.  The radiation coupling, Eq.~(\ref{eq:Hrf}), does not provide coupling between any of the four lowest-energy states and any of the higher states, justifying truncating our system to consist of only the four states.

A one-qubit operation changes the state of a single qubit, independent of the state of the other.  The field $B_1$ ($B_2$) will achieve this for qubit 1 (2), provided that $S_{1z,\downarrow\downarrow,\downarrow\uparrow}=S_{1z,\uparrow\downarrow,\uparrow\uparrow}$ ($S_{2z,\downarrow\downarrow,\uparrow\downarrow}=S_{2z,\downarrow\uparrow,\uparrow\uparrow}$).  For $J_{zz}=0$, this condition is nearly perfectly fulfilled, with a second-order error of $S_{1z,\downarrow\downarrow,\downarrow\uparrow}-S_{1z,\uparrow\downarrow,\uparrow\uparrow}\sim\frac{J_\perp^2}{\bar D \bar E}$.  The effect of $J_{zz}$ is more severe, resulting in an error $\sim \frac{J_{zz}}{\Delta E}$.  Thus, it is desirable to minimize $J_{zz}$ as much as possible.  This may be achievable through chemical engineering of the supramolecule to arrange the relative orientation of the easy axes into a configuration that results in a very small $J_{zz}$, akin to a ``magic angle'' effect.  Alternatively, one may use a switchable linker in the dimer to turn off the exchange coupling during the one-qubit operations \cite{chiesa_molecular_2015, ferrando-soria_modular_2016}.  Such an approach requires a fast, local probe to switch the linker state and the ability to measure the state of an individual supramolecule via spin resonance techniques.  In contrast, always-on coupling, while potentially leading to single-qubit errors, permit ensemble measurements, a less technically challenging approach.  

We perform simulations using the established Hamiltonian parameters for the (1) green and (2) purple variants of $\crmn$: $D_1=21$ GHz, $D_2=16.5$ GHz, $E_1=1.9$ GHz, $E_2=2.6$ GHz~\cite{ardavan_will_2007,garlatti_detailed_2014}.  In addition, we take $J_\perp=100$ MHz and $g=2$, while choosing different values of $J_{zz}$ as discussed below.  $J_\perp$ and $J_{zz}$ can be controlled during synthesis~\cite{timco_engineering_2009,ferrando-soria_modular_2016,chiesa_molecular_2015}. A basic one-qubit gate is a $\pi/2$ rotation, implemented by setting either $B_1$ or $B_2$ to 10~G for a sufficient time ($\sim18$~ns).  Different kinds of rotations ($X_i, Y_i,\ldots$) are achieved by setting the phase $\phi_i$ of the corresponding radiation field.  (The gate ``direction", e.g. $X$, does not correspond to a physical principal axis, e.g. $x$.)  We characterize the gate performance by applying it to the 20 states comprising  all the mutually unbiased bases of a four-state system~\cite{woottersOptimalStatedeterminationMutually1989}, determining the fidelity $\mathscr{F}=\left|\!\braket{\phi|\psi}\!\right|^2$ from the simulated ($\ket{\psi}$) and ideal ($\ket{\phi}$) output states for each input state and then averaging the 20 fidelities.  For $J_{zz}=0$, we obtain an average one-qubit gate fidelity $\bar{\mathscr{F}}=99.98\%$.  With  $J_{zz}=50$~MHz, the fidelity drops to $\bar{\mathscr{F}}=99.92\%$ while for $J_{zz}=100$~MHz, the fidelity is reduced to $\bar{\mathscr{F}}=99.7\%$, illustrating the importance of $J_{zz}$ in the error of the single-qubit gates.

Implementing two-qubit gates follows protocols developed for NMR-based quantum computing~\cite{nielsen_quantum_2010,vandersypen_nmr_2005}.  Such gates rely on the $J_\perp$ coupling  to entangle the states of the two qubits. To demonstrate a CNOT gate,
we follow a standard implementation protocol represented by $U_{\rm{CNOT}}=X_1[\bar Y_1 X_2][\bar X_1\bar Y_2] U_J(t_{\pi/2})Y_2$ (ignoring irrelevant phase factors), where $X_i$, etc., indicate $\pi/2$ rotations about the given axis for the \textit{i}th spin. 
Pairs of single-qubit gates enclosed in square brackets can be implemented simultaneously using two-tone pulses~\cite{barendsSuperconductingQuantumCircuits2014,piasecki_bimodal_1996}.
The process denoted $U_J(t_{\pi/2})$ indicates a period of free evolution ($B_1=B_2=0$) that entangles the states of the two qubits.  
The duration of this process is $t_{\pi/2}=\pi/2\delta=924$~ns, for the parameters of our simulations.  Small adjustments in the timing of each gate are made to optimize the performance of the CNOT.  

\begin{figure}[ht!]
\includegraphics[width=.9\linewidth]{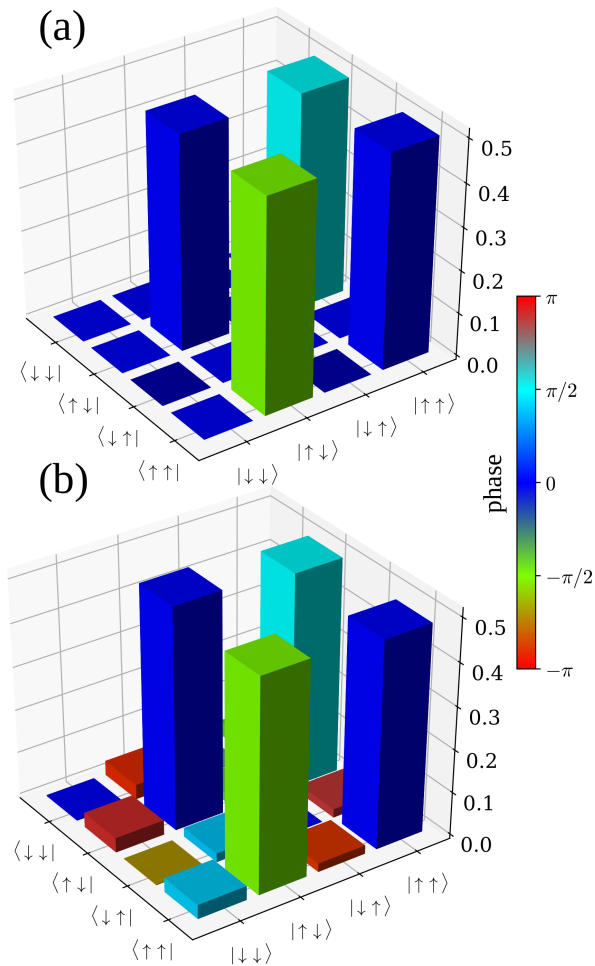}
\caption{\label{fig:cnotplot}Density matrices for a CNOT gate applied to the test input state $(\ket{\uparrow\uparrow}+i\ket{\downarrow\downarrow})\sqrt{2}$. The vertical axis represents the amplitude and the color represents the phase. (a) The ideal final state. (b) The simulated final state using $J_{zz}=50$ MHz. For this example, the simulation yields a calculated fidelity of 99.65\%.}
\end{figure}
Simulating the CNOT gate and evaluating average fidelity as described above yields the following results.  For $J_{zz}=0$, we obtain $\bar{\mathscr{F}}=99.94\%$, for  $J_{zz}=50$~MHz, $\bar{\mathscr{F}}=99.8\%$, and for $J_{zz}=100$~MHz,  $\bar{\mathscr{F}}=99.4\%$.  The reduction in fidelity with increasing $J_{zz}$ is almost entirely attributable to the  accumulated errors from single-qubit gates.   Figure \ref{fig:cnotplot} shows a comparison between the density matrices for a simulated CNOT gate and that of an ideal gate using a representative input state for $J_{zz}=50$~MHz.  


\begin{figure}[ht!]
\includegraphics[width=.9\linewidth]{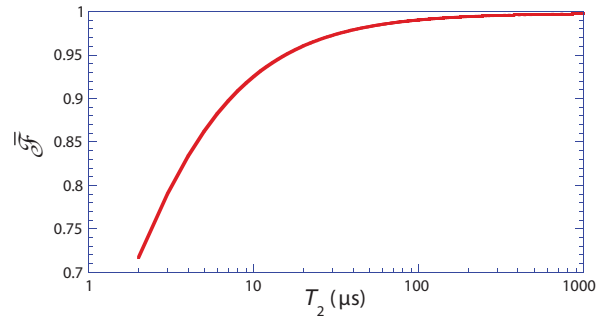}
\caption{\label{fig:dissplot} $T_2$ dependence of the average CNOT fidelity using $J_{zz}=50$~MHz on a semilog plot.}
\end{figure}

The results presented above do not include any effects of decoherence. 
To include the effects of decoherence, we adopt a model of ``pure dephasing" given that $T_1>>T_2$ in this system~\cite{ardavan_will_2007} and define a Lindblad collapse operator
\begin{equation}
  L_i=\frac{\sqrt{\gamma_i}}{2}\sigma_{zi}. \label{eq:lindblad}
\end{equation}
 We simulate the time evolution of our system using the Lindblad master equation:
\begin{equation}
  \dot{\rho}=-\frac{i}{h}[\mathscr{H}_{\rm{int}},\rho(t)]+\sum_i\Big[2L_i\rho(t)L_i^\dagger-\rho(t)L_i^\dagger L_i-L_i^\dagger L_i\rho(t)\Big]. \label{eq:meq}
\end{equation}
Numerically solving this equation using our optimized CNOT pulse sequence and $\gamma_1=\gamma_2=1/T_2$, allows us to compute fidelity $\mathscr{F}=\braket{\phi|\rho|\phi}$. Figure \ref{fig:dissplot} shows $\bar{\mathscr{F}}$ as a function of $T_2$. 
Near zero field we expect that decoherence in a $\crmn$ dimer due to field fluctuations will be suppressed because all of the transitions are clock transitions, leading to an increase in $T_2$.  The results in Fig.~\ref{fig:dissplot} show that an order of magnitude increase in $T_2$, like the increase reported in the Ho-based MNM~\cite{shiddiq_enhancing_2016}, can lead to a substantial enhancement in the CNOT gate fidelity.  Since $T_2$ for the molecular Cr$_7$Ni rings has been found to be $\sim$15~$\mu$s under optimized conditions~\cite{wedge_chemical_2012}, use of clock transitions to enhance the coherence further could appreciably impact gate fidelities in $\crmn$ heterodimers.

These results are encouraging for implementing quantum computing protocols in realistic supramolecular systems, such as those that have already been synthesized and characterized. All the necessary pulses can be readily implemented using existing ESR techniques.  The scheme presented can be extended to larger supramolecular structures with multiple molecular qubits, such as chains~\cite{whiteheadRingsThreadsLinkers2013}.  Such larger supramolecules, with or without switchable linkers, could then be used to implement quantum computing algorithms as has been done in NMR~\cite{vandersypen_nmr_2005} but at the much higher frequency scale of ESR, where near pure states can be achieved at milliKelvin temperatures and thus the scaling drawbacks of NMR quantum computing can be overcome.  The complicating effects of multiple couplings between qubits in the chain or structure can be ameliorated using refocusing pulse techniques, again borrowing from those developed for NMR quantum computing. Thus, by using clock transitions to enhance coherence, one should in principle be able to implement basic quantum computing protocols in supramolecular structures with good fidelity.  

\begin{acknowledgments}
We thank M. Foss-Feig for useful conversations.  Support for this work was provided by the U.S.~National Science Foundation under Grant Nos.~DMR-1310135 and DMR-1708692, and by the SUMO project of the QuantERA H2020 call, cofunded by Italian MIUR. J.R.F.~acknowledges the support of the Amherst College Senior Sabbatical Fellowship Program, funded in part by the H.~Axel Schupf '57 Fund for Intellectual Life.
\end{acknowledgments}

\bibliographystyle{apsrev4-1}
\bibliography{SMMrefs1}

\end{document}